\documentclass{ws-p8-50x6-00}

\newcommand{\zaa}{{\it Astron. Astrophys.}}

\newcommand{\zapjs}{{\it Astrophys.~J.~Sup.}}
\newcommand{\znp}{{\it Nucl.~Phys.}}

\newcommand{\zprc}{{\it Phys.~Rev.~C}}

\newcommand{\zADNDT}{{\it Atomic Data and Nuclear Data Tables}}

\newcommand{\sept}{$^7Li$}

\begin{document}

\title{Big-bang nucleosynthesis with the NACRE compilation}

\author{Alain Coc}

\address{CSNSM, B\^at. 104, 91405 Orsay Campus, France}

\author{Elisabeth Vangioni-Flam}

\address{IAP, 98$^{bis}$ Bd Arago 75014 Paris, France}  

\author{Michel Cass\'e}

\address{SAp, DAPNIA, DSM, CEA, 91191 Gif sur Yvette CEDEX
France, and IAP}


\maketitle

\abstracts{
We update the Big Bang Nucleosynthesis (BBN) calculations on the basis of
the recent NACRE compilation of reaction rates.
In particular, we calculate the uncertainties related to the nuclear reaction
rates on the abundances of \sept\ and compare our results with an other recent 
analysis.
}

Observations of light isotopes have flourished and in particular, high
quality $Li$ observations in the halo stars have been accumulated.
Thus, it is timely to reassess the determination of the baryonic density of 
the Universe in the light of advances in nuclear physics and astronomical 
observations. (For details see K.~Olive, these proceedings.)

Most of the important reactions for \sept\ production are available in
the NACRE compilation of thermonuclear reaction rates\cite{NACRE}
(others in our BBN network are adapted from earlier
compilations\cite{CF88,Smi93} or more recent works\cite{Bru99,Che99}).
One of the main innovative feature of NACRE with respect to former
compilations\cite{CF88} is that  uncertainties are analyzed in detail and
realistic lower and upper bounds for the rates are provided.
Using these low and high rate limits, it is thus possible to
calculate the effect of nuclear uncertainties on the light element yields.
In previous works, we have studied the influence of individual
reactions\cite{Van00b} or extreme yield limits\cite{Van00c}. 
Here we present results of Monte-Carlo calculations that allow to derive
statistical limits on the yields. Other such calculations have
been performed recently, based on a different compilation and analysis of
nuclear data (Nollett \& Burles\cite{Nol00}, hereafter NB).
The NB and NACRE compilations differ on several aspects.
The NB compilation addresses primordial nucleosynthesis while the NACRE one
is a general purpose compilation. Accordingly, the NB one contains a few
more reaction of interest to BBN and a few more data in the energy range of
interest. Also, {\em from the statistical point of view}, the rate
uncertainties are better defined in NB. However, in NB, the astrophysical
S--factors are fitted by splines which have no physical justification and can
produce local artifacts by following to closely experimental data points.
On the contrary, the NACRE compilation span wider energy ranges, and over
these ranges, the S--factors are fitted to functions reflecting theoretical
assumptions. This eliminates spurious local effects induced by a few erratic
data points associated with experimental problems rather than genuine
physical effect.
Due to the difficulty of defining a universal
statistical method for the wide set of reactions (each with its
peculiarities) and range of temperature, the NACRE rate limits correspond
to upper and lower bound rather than standard deviations.
In a previous study\cite{Van00c}, we considered all possible combinations of
these high and low rates and derived extreme upper and lower limits for
the \sept\ yields (Figure~1).
Here, instead, we ran Monte-Carlo calculations assuming a log--uniform
distribution for the rates between the low and high NACRE limits (keeping
the mean rates equal to the recommended rates).
For each $\eta$ value, we calculated the mean value and standard deviation
($\sigma$) of the \sept\ yield distribution. The corresponding 
2--$\sigma$ limits are represented in Figure~1 together with the results 
obtained by Burles et al.\cite{Bur00} with the NB compilation.
The results obtained from these two {\em independent} compilations agree 
very well reinforcing the confidence in these analyses. 

\begin{figure}[h]
\epsfxsize=30pc 
\epsfbox{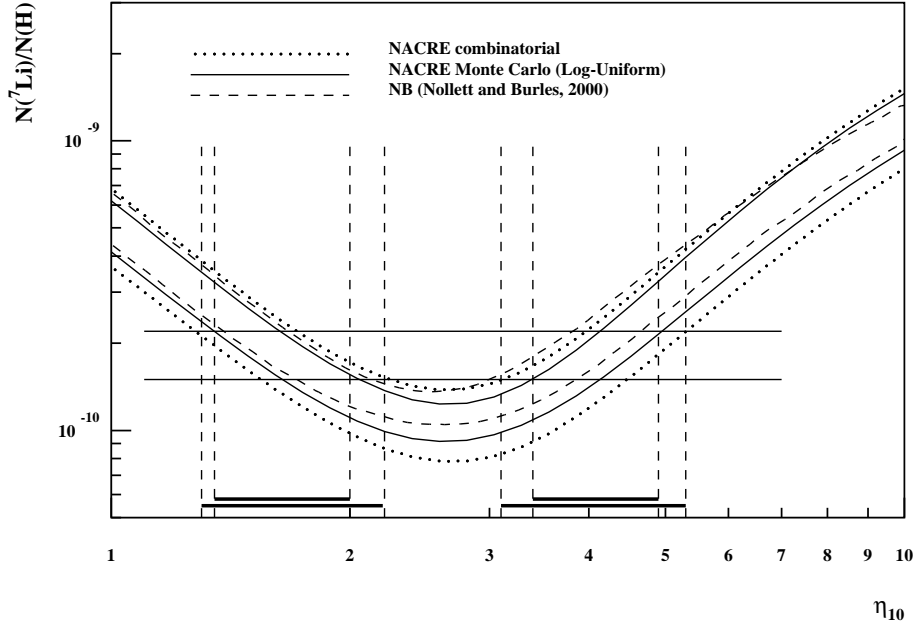} 
\caption{\sept\ limits obtained by combining NACRE limits\protect\cite{Van00c},
or Monte-Carlo calculations with the NACRE\protect\cite{NACRE} or
NB\protect\cite{Nol00,Bur00} compilations. Horizontal lines represent the
\sept\ observational limits while vertical line represent the deduced
$\eta_{10}$ limits.}
\end{figure}

The determination of $Li$ in halo stars indicate that
the Spite plateau is exceptionally thin ($<$ 0.05 dex).
We adopt the following range: 1.5 $10{^{-10}}$ $<$ ${^7}Li/H$ $<$ 2.2
$10{^{-10}}$, taking into account a maximum destruction of 0.1 dex.
Due to the valley shape of the $Li$ curve, we obtain two possible $\eta$ ranges
which are consistent respectively with a high $D$ and low ${^4}He$ observed
values, and with a low $D$ and high ${^4}He$ observed values.
As presently, none of these solutions can be excluded,
two possible ranges emerge: {\it i)} $1.35<\eta_{10}< 2.0$ corresponding
to $0.0049<\Omega_Bh^2<0.0073$ and {\it ii)} $3.4<\eta_{10}<4.9$ corresponding
to $0.012<\Omega_Bh^2<0.018$.
It is worth noting that these results (and also the extreme upper
limit\cite{Van00c} of $\eta_{10}<5.3$ or $\Omega_Bh^2<0.02$) are hardly
compatible with the recent MAXIMA and BOOMERANG baloons CMB
observations\cite{Jaf00} ($\Omega_Bh^2=0.032\pm0.005$).
However, other observations with the Cosmic Background Imager\cite{Pad01}
(CBI, Chile) lead to a lower value of $\Omega_Bh^2=0.009$ closer to the high 
$D$, low $\Omega_B$ range.
Clarification will hopefully come from future space missions 
(i.e. MAP and PLANCK Surveyor).

\end{document}